\documentstyle[amstex,amsfonts,preprint,aps]{revtex}
\tightenlines

\begin{document}
\title{Space time and rotations}
\author{A. Tartaglia}
\address{Gravity Probe B, Hansen Experimental Physics Labs, Stanford University,\\
Stanford (CA) and INFN, Torino, Italy\thanks{%
*Permanent address: Dip. Fisica, Politecnico, Torino, Italy.}}

\begin{abstract}
The paper considers the problem of finding the metric of space time around a
rotating, weakly gravitating body. Both external and internal metric tensors
are consistently found, together with an appropriate source tensor. All
tensors are calculated at the lowest meaningful approximation in a power
series. The two physical parameters entering the equations (the mass and the
angular momentum per unit mass) are assumed to be such that the mass effects
are negligible with respect to the rotation effects. A non zero Riemann
tensor is obtained. The order of magnitude of the effects at the laboratory
scale is such as to allow for experimental verification of the theory.
\end{abstract}

\date{\today }
\maketitle
\pacs{04.25.Nx, 04.90.+e}

\section{Introduction}

Rotational motion has a peculiarity on its own since it appears to be
absolute, unlike translational motion, which is purely relative. This
absoluteness of rotation posed a principle problem since the very time of
Newton with his rotating bucket example, which led precisely to the
conclusion that rotational motion was absolute\cite{newton}. A couple of
decades after Newton's Principia, George Berkeley questioned his notion of
an absolute space\cite{sciama} and successively Ernst Mach, looking for the
origin of inertia, stuck to the idea that even rotations are relative\cite
{mach}. Mach's approach was one of the ideas that inspired Einstein in
developing the general theory of relativity\cite{mach1}, although the
incorporation of Mach's principle into the theory is not entirely
satisfactory. In general, rotation is rather poorly treated in general
relativity. Its effect is essentially reduced to affecting the dynamical
mass of the rotating body, exactly as the translational motion does, and the
space time geometry through gravitomagnetic effects\cite{gravimag}, mediated
again by the mass of the source. This is the case of the Lense-Thirring
effect\cite{lense} and of the gravitomagnetic clock effects\cite{mashtart}.
However the treatment of the translational motion incorporates the very
notion of inertial reference frames and inertial observers, whereas rotating
systems do not identify any class of equivalent observers. The rotation
state itself is usually identified with respect to the asymptotic flatness
of space time, or more specifically to the constraint that the metric tensor
far away from the source assumes the Minkowski form. The rotation is
confronted with a non rotating (Minkowskian) space time, which serves as an
(absolute) reference frame.

The study of exact vacuum solutions of the Einstein equations has produced
many metric tensors, which are stationary and endowed with axial symmetry%
\cite{exact}. These solutions include situations which correspond to
rotating sources in an asymptotically flat space time. The most renowned is
of course the Kerr metric\cite{misner}, which contains two independent
parameters characterizing the source: the asymptotic mass and the asymptotic
angular momentum per unit mass. The latter quantity is interesting because
it actually does not depend on the mass itself, rather expresses a sort of
purely rotational property. Unfortunately however up to now it proved
impossible to link the vacuum Kerr solution to a satisfying internal metric
of a given matter distribution.

From general relativity we know that a mass curves space time around. If the
mass rotates, the peculiar motion introduces further warps in space time.
Suppose now that the influence of the very mass becomes weak (as it is
normally the case within the Solar system, for instance), but the rotation
stays important: would there be a residual effect on space time? In fact the
condition in which mass is negligible and rotation is not is rather easy to
obtain\cite{eurlet}.

This is the problem this paper will address. The first step will be to
identify the physical parameters describing the body and its rotation state.
Then the metric tensor outside the spinning body will be found, in the form
of an inverse powers of the distance development. Next will come the metric
tensor inside the rotating matter distribution and finally the source tensor
corresponding to the internal metric tensor and generating the external one.
As we shall see, a complete consistent solution will be found, producing
expectedly measurable effects on space time.

\section{General form of the approximated external metric tensor}

We start from the empty space-time line element in polar coordinates:
\[
ds^{2}=c^{2}dt^{2}-dr^{2}-r^{2}d\theta ^{2}-r^{2}\sin ^{2}\theta d\phi ^{2}
\]

Let us assume that the presence of an axially symmetric steadily rotating
mass at the origin introduces a perturbation such that the metric becomes
\begin{equation}
ds^{2}=c^{2}\left( 1+{\frak h}_{00}\right) dt^{2}-\left( 1+{\frak h}%
_{rr}\right) dr^{2}-r^{2}\left( 1+{\frak h}_{\theta \theta }\right) d\theta
^{2}-r^{2}\sin ^{2}\theta \left( 1+{\frak h}_{\phi \phi }\right) d\phi ^{2}+2%
{\frak h}_{t\phi }r\sin \theta cdtd\phi  \label{ap}
\end{equation}
with the condition
\[
{\frak h}_{\mu \nu }<<1
\]

The elements of the perturbation tensor, because of the assumed symmetries,
depend geometrically on $r$ and $\theta $ only. Physically the perturbation
must depend on two quantities, which characterize the body and its motion:
the mass $M$ and the angular velocity $\Omega $, or its combination with the
mass into the angular momentum $J$. Both $\Omega $ and $J$ are defined from
the viewpoint of an inertial observer at rest with respect to the rotation
axis of the body, thus coinciding with the components of the corresponding
three-vectors on that axis. Expressing the mass and the angular momentum as
lengths it is possible to introduce the quantities $\mu =GM/c^{2}$ and $%
a=J/Mc$; $a$, apart from a factor depending on the shape of the body and the
matter distribution inside it, is proportional to $\Omega $ and does not
depend on $M$ any more.

Once the reference frame has been fixed, the sign of the parameter $a$
varies according to the two directions of rotation: $a$ is odd for reversal
of time. The line element (\ref{ap}) must of course be even in time; this
implies that the diagonal terms of the metric tensor must be even in time
too, i.e. must contain even powers of $a$ only. In order to have the mixed
term in the line element being even in time, since it contains $dt$, we must
impose to the off diagonal term of the metric tensor to be odd, which means
to contain odd powers of $a$ only and no pure powers of $\mu $.

In practice, introducing the dimensionless variables $\varepsilon =\mu /r$
and $\alpha =a/r$ and expressing the $r$ dependence in the form of an
inverse powers development leads to:
\begin{eqnarray*}
{\frak h}_{00} &=&A_{0}\varepsilon +B_{0}\alpha ^{2}+... \\
{\frak h}_{rr} &=&A_{1}\varepsilon +B_{1}\alpha ^{2}+... \\
{\frak h}_{\theta \theta } &=&A_{2}\varepsilon +B_{2}\alpha ^{2}+... \\
{\frak h}_{\phi \phi } &=&A_{3}\varepsilon +B_{3}\alpha ^{2}+... \\
{\frak h}_{t\phi } &=&A_{4}\alpha +B_{4}\varepsilon \alpha +...
\end{eqnarray*}
The $A$ and $B$ coefficients are functions of $\theta $ only, which in turn
is the physical angle between the position three-vector and the angular
velocity axial three-vector.

Considering ${\frak h}_{t\phi }$ we see that the linear term (in $\alpha $),
when introduced in (\ref{ap}), produces a constant, which means that the
metric would not be flat at infinity. To avoid this it must be $A_{4}=0$.
Finally, considering and comparing the orders of magnitude of the different
terms, we assume that, in general, the mass contributions ($\varepsilon $
terms) are negligible with respect to the rest; as said in the introduction
this condition is rather simple to obtain\cite{eurlet}. Under this
assumption the approximated line element will be
\[
ds^{2}=c^{2}\left( 1+B_{0}\alpha ^{2}\right) dt^{2}-\left( 1+B_{1}\alpha
^{2}\right) dr^{2}-r^{2}\left( 1+B_{2}\alpha ^{2}\right) d\theta
^{2}-r^{2}\sin ^{2}\theta \left( 1+B_{3}\alpha ^{2}\right) d\phi ^{2}
\]
or, more explicitly,
\begin{equation}
ds^{2}=c^{2}\left( 1+B_{0}\frac{a^{2}}{r^{2}}\right) dt^{2}-\left( 1+B_{1}%
\frac{a^{2}}{r^{2}}\right) dr^{2}-r^{2}\left( 1+B_{2}\frac{a^{2}}{r^{2}}%
\right) d\theta ^{2}-r^{2}\sin ^{2}\theta \left( 1+B_{3}\frac{a^{2}}{r^{2}}%
\right) d\phi ^{2}  \label{line}
\end{equation}

\subsection{Conditions to be imposed on the metric tensor}

From the metric tensor corresponding to (\ref{line}) one can calculate the
Ricci tensor up to terms in $a^{2}$. Since, by hypothesis, we are in empty
space time, all the elements of the Ricci tensor must vanish. This condition
corresponds to the following equations for the $B$'s (a $^{\prime }$ denotes
differentiation with respect to $\theta $):
\begin{eqnarray*}
\left( 2B_{0}^{\prime \prime }+4B_{0}\right) \sin \theta +2B_{0}^{\prime
}\cos \theta  &=&0 \\
\left( 4B_{3}^{\prime }+2B_{1}^{\prime }+6B_{0}^{\prime }\right) \sin \theta
+4\left( B_{3}-B_{2}\right) \cos \theta  &=&0 \\
\left( 4B_{3}+4B_{2}+2B_{1}^{\prime \prime }+8B_{1}+12B_{0}\right) \sin
\theta +2B_{1}^{\prime }\cos \theta  &=&0 \\
\left( 2B_{3}^{\prime \prime }-4B_{3}+4B_{2}+2B_{1}^{\prime \prime
}+2B_{0}^{\prime \prime }-4B_{0}\right) \sin \theta +2\left( 2B_{3}^{\prime
}-B_{2}^{\prime }\right) \cos \theta  &=&0 \\
\left( 2B_{3}^{\prime \prime }-4B_{0}\right) \sin \theta +\left(
4B_{3}^{\prime }-2B_{2}^{\prime }+2B_{1}^{\prime }+2B_{0}^{\prime }\right)
\cos \theta  &=&0
\end{eqnarray*}

Only four out of these equations can be independent since the Ricci tensor
is symmetric and consequently it can be diagonalized at any moment and place
in space time. The number of independent equations is further reduced
because of the rotation symmetry of the body and steadiness of the motion.
Solving for the $B$ functions one obtains:

\begin{align*}
B_{0}& =C_{0}\cos \theta +D_{0}\left( 1+\frac{1}{2}\cos \theta \ln \left(
\frac{1-\cos \theta }{1+\cos \theta }\right) \right) \\
B_{1}& =-2f\cos ^{2}\theta -3C_{0}\cos \theta +C_{1}\cos ^{2}\theta \\
B_{2}& =2f\sin ^{2}\theta -C_{1}\sin ^{2}\theta -4\left( \sin \theta \cos
\theta \right) f^{\prime }+\allowbreak \left( \cos ^{2}\theta \right)
f^{\prime \prime } \\
B_{3}& =\frac{\cos ^{3}\theta }{\sin \theta }f^{\prime }
\end{align*}
where $f=f\left( \theta \right) $ is an arbitrary function. $C_{0}$, $C_{1}$%
and $D_{0}$ are constants. Actually only finite solutions can be accepted
(in order the development to be consistent all the $B$ functions must be $%
\sim 1$), consequently it must be $D_{0}=0$.

\begin{align}
B_{0}& =C_{0}\cos \theta  \nonumber \\
B_{1}& =-2f\cos ^{2}\theta -3C_{0}\cos \theta +C_{1}\cos ^{2}\theta
\nonumber \\
B_{2}& =2f\sin ^{2}\theta -C_{1}\sin ^{2}\theta -4\left( \sin \theta \cos
\theta \right) f^{\prime }+\allowbreak \left( \cos ^{2}\theta \right)
f^{\prime \prime }  \label{bees} \\
B_{3}& =\frac{\cos ^{3}\theta }{\sin \theta }f^{\prime }  \nonumber
\end{align}

If it is $C_{0}\neq 0$ the correction to the $g_{00}$ term of the metric has
the form of a dipolar potential, consistent with the axial character of the
angular velocity three-vector.

Now we can calculate the non zero terms of the Riemann tensor, which are
\[
\begin{array}{ll}
R_{rtr}^{t}=-3C_{0}\frac{a^{2}}{r^{4}}\cos \theta & R_{rt\theta }^{t}=-\frac{%
3}{2}C_{0}\frac{a^{2}}{r^{3}}\sin \theta \\
R_{\theta t\theta }^{t}=\frac{3}{2}C_{0}\frac{a^{2}}{r^{2}}\cos \theta &
R_{\phi t\phi }^{t}=\frac{3}{2}C_{0}\frac{a^{2}}{r^{2}}\cos \theta \sin
^{2}\theta \\
R_{\theta r\theta }^{r}=\frac{3}{2}C_{0}\frac{a^{2}}{r^{2}}\cos \theta &
R_{\phi r\phi }^{r}=\frac{3}{2}C_{0}\frac{a^{2}}{r^{2}}\cos \theta \sin
^{2}\theta \\
R_{\phi \theta \phi }^{r}=\frac{3}{2}C_{0}\frac{a^{2}}{r}\sin ^{3}\theta &
R_{\phi \theta \phi }^{\theta }=-3C_{0}\frac{a^{2}}{r^{2}}\cos \theta \sin
^{2}\theta
\end{array}
\]
The presence of these terms indicates the existence of real physical effects
depending solely on the rotation of the body. As it is seen they are there
only when $C_{0}\neq 0$.

Though in the form of a power series the metric corresponding to (\ref{bees}%
) belongs to Weyl's class of axially symmetric stationary vacuum solutions
of the Einstein field equations\cite{exact}.

\section{Determining the source tensor}

In order to find the source tensor for the metric corresponding to (\ref
{bees}) it is convenient to proceed as it is usually done in the linearized
theory of gravity, writing the metric tensor as the sum of the Minkowski
metric tensor (written in polar coordinates, in our case) $g_{_{flat}\mu \nu
}$ and a small correcting tensor $h_{\mu \nu }$.
\[
g_{\mu \nu }=g_{_{flat}\mu \nu }+h_{\mu \nu }
\]
The explicit expressions for the perturbations are
\[
h_{\mu \nu }=\left(
\begin{array}{cccc}
B_{0}\frac{a^{2}}{r^{2}} & 0 & 0 & 0 \\
0 & -B_{1}\frac{a^{2}}{r^{2}} & 0 & 0 \\
0 & 0 & -B_{2}a^{2} & 0 \\
0 & 0 & 0 & -B_{3}a^{2}\sin ^{2}\theta
\end{array}
\right)
\]

The trace $h=h_{\mu}^{\mu}$ is

\[
h=\frac{a^{2}}{r^{2}}\left( B_{0}+B_{1}+B_{2}+B_{3}\right)
\]
Following the standard method, let us define the auxiliary tensor $\overline{%
h}_{\mu \nu }=h_{\mu \nu }-\frac{1}{2}g_{_{flat}\mu \nu }h$. It is
\begin{eqnarray}
\overline{h}_{0}^{0} &=&\frac{1}{2}\frac{a^{2}}{r^{2}}\left(
B_{0}-B_{1}-B_{2}-B_{3}\right)   \nonumber \\
\overline{h}_{r}^{r} &=&-\frac{1}{2}\frac{a^{2}}{r^{2}}\left(
B_{0}-B_{1}+B_{2}+B_{3}\right)   \label{hbari} \\
\overline{h}_{\theta }^{\theta } &=&-\frac{1}{2}\frac{a^{2}}{r^{2}}\left(
B_{0}+B_{1}-B_{2}+B_{3}\right)   \nonumber \\
\overline{h}_{\phi }^{\phi } &=&-\frac{1}{2}\frac{a^{2}}{r^{2}}\left(
B_{0}+B_{1}+B_{2}-B_{3}\right)   \nonumber
\end{eqnarray}

The determinant of the metric tensor will be assumed to coincide with the
one of flat space time, since all corrections are of order $a^{4}$: $%
g=g_{_{flat}}=-c^{2}r^{4}\sin ^{2}\theta $.

Under the hypothesis of a weak field and considering that the metric tensor
is stationary, one can write
\[
\nabla ^{2}\overline{h}_{\mu }^{\nu }=16\pi S_{\mu }^{\nu }
\]
whence
\[
\overline{h}_{\mu }^{\nu }=-4\int \frac{S_{\mu }^{\nu }}{\left| \overline{r}-%
\overline{r}^{\prime }\right| }\sqrt{\gamma }d^{3}x^{\prime }
\]
The square root of the determinant of the space part of the Minkowski metric
tensor is written as $\sqrt{\gamma }$. Developing in multipolar components
and keeping the lowest order contributions, one has
\[
\overline{h}_{\mu }^{\nu }=-\frac{4}{r}\int S_{\mu }^{\nu }\sqrt{\gamma }%
d^{3}x^{\prime }-\frac{4\cos \theta }{r^{2}}\int S_{\mu }^{\nu }r^{\prime
}\cos \theta ^{\prime }\sqrt{\gamma }d^{3}x^{\prime }
\]
Looking at the (\ref{hbari}) we see that there is no $1/r$ term. This means
that $S_{\mu }^{\nu }$ must be odd with respect to an odd number of
integration variables. Considering the independence from $\phi $ (invariance
for rotation) and the nature of $r$, only $\theta $ is left. Let us assume
that
\[
S_{\mu \nu }={\cal S}_{\mu \nu }w\left( \theta \right)
\]
with ${\cal S}_{\mu \nu }$ depending at most on $r$ and $w\left( \theta
\right) $ an odd function of $\theta $ (odd with respect to the equatorial
plane of the body), so that in general
\begin{eqnarray}
\overline{h}_{\mu }^{\nu } &=&-\frac{4\cos \theta }{r^{2}}\int_{0}^{\pi
}\int_{0}^{{\cal R}}\int_{0}^{2\pi }{\cal S}_{\mu }^{\nu }r^{\prime
3}w\left( \theta ^{\prime }\right) \cos \theta ^{\prime }\sin \theta
^{\prime }dr^{\prime }d\theta ^{\prime }d\phi ^{\prime }  \nonumber \\
&=&-8\pi \frac{\cos \theta }{r^{2}}\int_{0}^{\pi }\int_{0}^{{\cal R}}{\cal S}%
_{\mu }^{\nu }r^{\prime 3}w\left( \theta ^{\prime }\right) \cos \theta
^{\prime }\sin \theta ^{\prime }dr^{\prime }d\theta ^{\prime }
\label{general}
\end{eqnarray}

The result of the integration depends of course also on the shape of the
body and in particular on the shape of the meridian section of the body
expressed through the function $r^{\prime }={\cal R}\left( \theta ^{\prime
}\right) $ representing the border of that section. A further assumption may
be that the body is homogeneous so that ${\cal S}_{\mu }^{\nu }$ is not
depending on $r^{\prime }$ either (or we can refer to the average value of $%
{\cal S}_{\mu }^{\nu }$ along the radius). So one has
\begin{equation}
\overline{h}_{\mu }^{\nu }=-2\pi \frac{\cos \theta }{r^{2}}{\cal S}_{\mu
}^{\nu }\int_{0}^{\pi }{\cal R}^{4}\left( \theta ^{\prime }\right) w\left(
\theta ^{\prime }\right) \cos \theta ^{\prime }\sin \theta ^{\prime }d\theta
^{\prime }=-2\pi F\frac{\cos \theta }{r^{2}}{\cal S}_{\mu }^{\nu }
\label{forma}
\end{equation}
where $F$ is a constant depending on the shape of the section of the body.

Comparing this result with (\ref{hbari}) we see that it must be

\begin{eqnarray}
{\cal S}_{0}^{0} &=&-\frac{1}{4\pi }\frac{a^{2}}{F}\frac{\left(
B_{0}-B_{1}-B_{2}-B_{3}\right) }{\cos \theta }  \nonumber \\
{\cal S}_{r}^{r} &=&\frac{1}{4\pi }\frac{a^{2}}{F}\frac{\left(
B_{0}-B_{1}+B_{2}+B_{3}\right) }{\cos \theta }  \label{tmedio} \\
{\cal S}_{\theta }^{\theta } &=&\frac{1}{4\pi }\frac{a^{2}}{F}\frac{\left(
B_{0}+B_{1}-B_{2}+B_{3}\right) }{\cos \theta }  \nonumber \\
{\cal S}_{\phi }^{\phi } &=&\frac{1}{4\pi }\frac{a^{2}}{F}\frac{\left(
B_{0}+B_{1}+B_{2}-B_{3}\right) }{\cos \theta }  \nonumber
\end{eqnarray}
The left hand side does not depend on $\theta $, so of course the same
should happen to the right hand side. This fact poses constraints on the $f$
function appearing in (\ref{bees}). Equating all (\ref{tmedio}) to constants
one obtains that it must be
\[
f=\frac{W}{\cos \theta }+\frac{C_{1}}{2}
\]
Now $W$ is a constant. Consequently the $B$ functions are
\begin{align}
B_{0}& =C_{0}\cos \theta  \nonumber \\
B_{1}& =-\left( 2W+3C_{0}\right) \cos \theta  \label{beequasi} \\
B_{2}& =W\cos \theta  \nonumber \\
B_{3}& =W\cos \theta  \nonumber
\end{align}
and the source tensor is
\begin{eqnarray}
S_{0}^{0} &=&-\frac{1}{\pi }\frac{a^{2}}{F}C_{0}w\left( \theta \right)
\nonumber \\
S_{r}^{r} &=&\frac{1}{\pi }\frac{a^{2}}{F}\left( \allowbreak C_{0}+W\right)
w\left( \theta \right)  \nonumber \\
S_{\theta }^{\theta } &=&-\frac{1}{2\pi }\frac{a^{2}}{F}\left(
C_{0}+W\right) w\left( \theta \right)  \label{tquasi} \\
S_{\phi }^{\phi } &=&-\frac{1}{2\pi }\frac{a^{2}}{F}\left( C_{0}+W\right)
w\left( \theta \right)  \nonumber
\end{eqnarray}

Of course the source tensor must satisfy also the zero divergence condition.
Considering the symmetries and the fact that the weak field approximation
holds supposedly also inside the body (use of the flat space time
Christoffel symbols for the covariant derivatives), the null four-divergence
condition reduces to
\begin{eqnarray}
\frac{d\left( {\cal S}_{\theta }^{\theta }w\left( \theta \right) \right) }{%
d\theta }+\frac{\cos \theta }{\sin \theta }w\left( \theta \right) \left(
{\cal S}_{\theta }^{\theta }-{\cal S}_{\phi }^{\phi }\right)  &=&0
\label{divcond} \\
2{\cal S}_{r}^{r}-{\cal S}_{\theta }^{\theta }-{\cal S}_{\phi }^{\phi } &=&0
\nonumber
\end{eqnarray}
If $w(\theta )$ must be odd through the equatorial plane, eq.s (\ref{divcond}%
) can be satisfied only with ${\cal S}_{r}^{r}={\cal S}_{\theta }^{\theta }=%
{\cal S}_{\phi }^{\phi }=0$. This fact in turn implies that
\[
W=-C_{0}
\]
The general forms of the $B$'s and of the source tensor are then
\begin{align}
B_{0}& =C_{0}\cos \theta   \nonumber \\
B_{1}& =-C_{0}\cos \theta   \label{quasifin} \\
B_{2}& =-C_{0}\cos \theta   \nonumber \\
B_{3}& =-C_{0}\cos \theta   \nonumber
\end{align}
and
\begin{eqnarray}
S_{0}^{0} &=&-\frac{1}{\pi }\frac{a^{2}}{F}C_{0}w\left( \theta \right)
\nonumber \\
S_{r}^{r} &=&0  \label{tquasifin} \\
S_{\theta }^{\theta } &=&0  \nonumber \\
S_{\phi }^{\phi } &=&0  \nonumber
\end{eqnarray}

\section{The internal metric tensor}

The next step is to determine a consistent metric tensor inside the matter
distribution. As in the case of the external solution it is convenient to
use the physical dimensionless variables
\begin{eqnarray*}
\varepsilon  &=&\frac{Gm}{c^{2}r} \\
\alpha  &=&\frac{a}{r}
\end{eqnarray*}
where now both $m$ and $a$ depend on $r$. For an homogeneous, rigidly
rotating sphere it would be $m=\frac{4}{3}\pi \rho r^{3}$ and $a=\frac{2}{5}%
\frac{r^{2}}{c}\Omega $, where $\rho $ is the matter density and $\Omega $
is the angular velocity. In general $a$ would be expressed as a numerical
factor depending on the shape, size and kind of matter distribution
multiplying the angular velocity of the body and the square of some
characteristic distance from the axis. On the basis of these considerations
the dimensionless variables inside the body can be written as
\begin{eqnarray*}
\varepsilon  &=&\frac{G\rho }{c^{2}}r^{2} \\
\alpha  &=&\frac{\Omega r}{c}
\end{eqnarray*}
transferring all shape effects into the coefficients that will multiply the
variables.

Here again we assume that
\[
\varepsilon <<\alpha
\]

Now let us write the internal line element as a power expansion in $r$ where
the same general symmetries as in the external case should hold. In practice
the extra-diagonal term is negligible, as well as the mass terms.

The proposed line element is then
\begin{equation}
ds^{2}=c^{2}(1+\beta _{0}\frac{\Omega ^{2}}{c^{2}}r^{2})dt^{2}-\left(
1+\beta _{1}\frac{\Omega ^{2}}{c^{2}}r^{2}\right) dr^{2}-r^{2}\left( 1+\beta
_{2}\frac{\Omega ^{2}}{c^{2}}r^{2}\right) d\theta ^{2}-r^{2}\sin ^{2}\theta
\left( 1+\beta _{3}\frac{\Omega ^{2}}{c^{2}}r^{2}\right) d\phi ^{2}
\label{intlin}
\end{equation}
where the $\beta $'s are functions of $\theta $ only. From there one can
straightforwardly calculate the Einstein tensor
\begin{eqnarray*}
G_{t}^{t} &=&-\frac{\Omega ^{2}}{c^{2}}\frac{(\beta _{3}^{\prime \prime
}+8\beta _{3}+10\beta _{2}+\beta _{1}^{\prime \prime }-6\beta _{1})\sin
\theta +(2\beta _{3}^{\prime }-\beta _{2}^{\prime }+\beta _{1}^{\prime
})\cos \theta }{2\sin \theta } \\
G_{r}^{r} &=&-\frac{\Omega ^{2}}{c^{2}}\frac{(\beta _{3}^{\prime \prime
}+2\beta _{3}+4\beta _{2}-2\beta _{1}+\beta _{0}^{\prime \prime }+4\beta
_{0})\sin \theta +(2\beta _{3}^{\prime }-\beta _{2}^{\prime }+\beta
_{0}^{\prime })\cos \theta }{2\sin \theta } \\
G_{r}^{\theta } &=&\frac{\Omega ^{2}}{c^{2}}\frac{(2\beta _{3}^{\prime
}-\beta _{1}^{\prime }+\beta _{0}^{\prime })\sin \theta +2(\beta _{3}-\beta
_{2})\cos \theta }{2r\sin \theta } \\
G_{\theta }^{r} &=&r\frac{\Omega ^{2}}{c^{2}}\frac{(2\beta _{3}^{\prime
}-\beta _{1}^{\prime }+\beta _{0}^{\prime })\sin \theta +2(\beta _{3}-\beta
_{2})\cos \theta }{2\sin \theta } \\
G_{\theta }^{\theta } &=&-\frac{\Omega ^{2}}{c^{2}}\frac{(6\beta _{3}-2\beta
_{1}+4\beta _{0})\sin \theta +(\beta _{1}^{\prime }+\beta _{0}^{\prime
})\cos \theta }{2\sin \theta } \\
G_{\phi }^{\phi } &=&-\frac{\Omega ^{2}}{c^{2}}\frac{(6\beta _{2}+\beta
_{1}^{\prime \prime }-2\beta _{1}+\beta _{0}^{\prime \prime }+4\beta _{0})}{2%
}
\end{eqnarray*}

Imposing the four-divergence of $G_{\mu }^{\nu }$ to be zero produces the
two equations
\begin{eqnarray*}
\frac{1}{2}\beta _{1}^{\prime \prime }-\frac{1}{2}\beta _{0}^{\prime \prime
}-\beta _{3}^{\prime \prime }+\beta _{3}-\beta _{2}-\frac{\cos \theta }{\sin
\theta }\left( 2\beta _{3}^{\prime }-\beta _{2}^{\prime }+\frac{\beta
_{0}^{\prime }}{2}-\frac{\beta _{1}^{\prime }}{2}\right) &=&0 \\
\frac{1}{2}\beta _{1}^{\prime }-\frac{1}{2}\beta _{0}^{\prime }-\beta
_{3}^{\prime }-\allowbreak \frac{\cos \theta }{\sin \theta }\beta _{3}+\frac{%
\cos \theta }{\sin \theta }\beta _{2} &=&0
\end{eqnarray*}

These equations are not independent from each other since differentiating
the second one, then subtracting it from the first one, the second equation
is obtained again. A solution is found when $\beta =\beta _{0}=-\beta
_{1}=-\beta _{2}=-\beta _{3}$. This same solution brings all the $G_{i}^{j}$%
's to $0$ too.

The Einstein tensor must be proportional to (\ref{tquasifin}); let us call $%
\chi $ the proportionality constant. The equality condition reduces to
\[
-\frac{\Omega ^{2}}{c^{2}}\frac{(\beta _{3}^{\prime \prime }+8\beta
_{3}+10\beta _{2}+\beta _{1}^{\prime \prime }-6\beta _{1})\sin \theta
+(2\beta _{3}^{\prime }-\beta _{2}^{\prime }+\beta _{1}^{\prime })\cos
\theta }{2\sin \theta }=-\frac{1}{\pi }\chi \frac{a^{2}}{F}C_{0}w\left(
\theta \right)
\]
i.e.

\begin{equation}
\frac{(\beta ^{\prime \prime }+6\beta )\sin \theta +\beta ^{\prime }\cos
\theta }{\sin \theta }=\frac{1}{\pi }\chi \frac{a^{2}c^{2}}{F\Omega ^{2}}%
C_{0}w\left( \theta \right)  \label{final0}
\end{equation}
Eq. (\ref{final0}) may be rewritten as ($K=\frac{1}{\pi }\chi \frac{%
a^{2}c^{2}}{F\Omega ^{2}}C_{0}$)
\begin{equation}
\frac{d^{2}\beta }{d\theta ^{2}}+\frac{d\beta }{d\theta }\frac{\cos \theta }{%
\sin \theta }+6\beta =Kw\left( \theta \right)  \label{finale}
\end{equation}

\allowbreak Further conditions to be imposed are related to the continuity
of the metric tensor at the border of the rotating body. These conditions
reduce in practice to
\begin{equation}
\beta \frac{\Omega ^{2}}{c^{2}}{\cal R}^{2}=B_{0}\frac{a^{2}}{{\cal R}^{2}}
\label{continui}
\end{equation}
from where one has
\[
\beta =\frac{a^{2}c^{2}}{\Omega ^{2}{\cal R}^{4}}C_{0}\cos \theta
\]
Introducing this $\beta $ into (\ref{finale}) gives
\begin{equation}
\frac{1}{{\cal R}^{4}}\cos \theta +2\frac{{\cal R}^{\prime }}{{\cal R}^{5}}%
\sin \theta +5\frac{\left( {\cal R}^{\prime }\right) ^{2}}{{\cal R}^{6}}\cos
\theta -\frac{{\cal R}^{\prime \prime }}{{\cal R}^{5}}\cos \theta -\frac{%
{\cal R}^{\prime }}{{\cal R}^{5}}\frac{\cos ^{2}\theta }{\sin \theta }=\frac{%
1}{4\pi }\frac{\chi }{F}w\left( \theta \right)  \label{issima}
\end{equation}

Once ${\cal R}$ is chosen (\ref{issima}) gives the expression for $w\left(
\theta \right) $.

\subsection{The case of a rotating sphere}

Let us assume our rotating body is a solid homogeneous sphere. In this case
it is
\[
{\cal R}=R=\text{constant}
\]
and
\[
a=\frac{2}{5}\frac{\Omega }{c}R^{2}
\]

From (\ref{issima}) we see that
\[
w\left( \theta \right) =\cos \left( \theta \right)
\]
On the other hand one has from (\ref{forma})
\[
F=\frac{2}{3}R^{4}
\]
Then, again from (\ref{issima})
\begin{equation}
\chi =\frac{8}{3}\pi  \label{chi}
\end{equation}

Finally we can list the explicit expressions for the line elements and the
source tensor.

Internal line element:
\begin{eqnarray}
ds^{2} &=&c^{2}(1+\frac{4}{25}C_{0}\frac{\Omega ^{2}}{c^{2}}r^{2}\cos \theta
)dt^{2}-\left( 1-\frac{4}{25}C_{0}\frac{\Omega ^{2}}{c^{2}}r^{2}\cos \theta
\right) dr^{2}  \nonumber \\
&&-r^{2}\left( 1-\frac{4}{25}C_{0}\frac{\Omega ^{2}}{c^{2}}r^{2}\cos \theta
\right) d\theta ^{2}-r^{2}\sin ^{2}\theta \left( 1-\frac{4}{25}C_{0}\frac{%
\Omega ^{2}}{c^{2}}r^{2}\cos \theta \right) d\phi ^{2}  \label{linesint}
\end{eqnarray}

External line element
\begin{eqnarray}
ds^{2} &=&c^{2}\left( 1+\frac{4}{25}C_{0}\frac{\Omega ^{2}R^{4}}{c^{2}r^{2}}%
\cos \theta \right) dt^{2}-\left( 1-\frac{4}{25}C_{0}\frac{\Omega ^{2}R^{4}}{%
c^{2}r^{2}}\cos \theta \right) dr^{2}  \nonumber \\
&&-r^{2}\left( 1-\frac{4}{25}C_{0}\frac{\Omega ^{2}R^{4}}{c^{2}r^{2}}\cos
\theta \right) d\theta ^{2}-r^{2}\sin ^{2}\theta \left( 1-\frac{4}{25}C_{0}%
\frac{\Omega ^{2}R^{4}}{c^{2}r^{2}}\cos \theta \right) d\phi ^{2}
\label{linesex}
\end{eqnarray}

Source tensor
\begin{eqnarray}
S_{0}^{0} &=&-\frac{6}{25\pi }\frac{\Omega ^{2}}{c^{2}}C_{0}\cos \theta
\nonumber \\
S_{r}^{r} &=&0  \label{sources} \\
S_{\theta }^{\theta } &=&0  \nonumber \\
S_{\phi }^{\phi } &=&0  \nonumber
\end{eqnarray}

\section{Interpretation of the source tensor}

To interpret the tensor and to conjecture a value of $C_{0}$ let us consider
what happens to the line element in a weak field approximation inside a
homogeneous sphere when considering the pure effect of mass. In that case we
expect the $g_{00}$ element of the metric tensor to be corrected by the
local Newtonian potential in the form
\[
c^{2}h_{00}=-\frac{2}{r}\int G\rho \sqrt{\gamma }d^{3}x^{\prime }
\]
Suppose now that the same role is played by the centrifugal potential inside
the body. That potential at a given point is $\frac{1}{2}\Omega
^{2}r^{2}\sin ^{2}\theta =\frac{1}{2}\Omega ^{2}\frac{r^{3}}{r}\sin
^{2}\theta $. Suppose you want to write it in the form of a volume integral
up to a given $r$ and a given $\theta $; it could be:
\[
\frac{1}{2}\Omega ^{2}r^{2}\sin ^{2}\theta =\frac{3}{2\pi }\frac{\Omega ^{2}%
}{r}\int_{0}^{2\pi }\int_{0}^{\theta }\int_{0}^{r}\cos \theta ^{\prime }%
\sqrt{\gamma }d^{3}x^{\prime }=\Omega ^{2}r^{2}\int_{0}^{\theta }\cos \theta
^{\prime }\sin \theta ^{\prime }d\theta ^{\prime }
\]
In this way, recalling also the 'coupling constant' (\ref{chi}), the same
role as $G\rho /c^{2}$ before, is now played by
\[
-4\frac{\Omega ^{2}}{c^{2}}\cos \theta
\]
In the case of a pure mass effect the $T_{0}^{0}$ term of the stress energy
tensor would contain precisely $G\rho /c^{2}$. Continuing on the same line
of thought and looking at (\ref{sources}) we expect
\[
\frac{6}{25\pi }\frac{\Omega ^{2}}{c^{2}}C_{0}\cos \theta =4\frac{\Omega ^{2}%
}{c^{2}}\cos \theta
\]
that implies
\[
C_{0}=\frac{50}{3}\pi \sim 10^{2}
\]

\section{Zero mass limit}

Assuming a rigidly rotating body one must allow for some kind of force
keeping the whole thing together against centrifugal forces. This force,
when self gravitation is negligible, is provided by elasticity. To account
for it the elastic energy momentum tensor should be added to the source
tensor (\ref{tquasifin}). A simpler approach is to think of an elastic
membrane constraining the body and balancing the centrifugal push on the
surface. In that case there would be a pure surface tension. If furthermore
we assume that the membrane is indeed made of a big number of independent
rings than the stress is purely directed along the 'parallels' of the
membrane. Its value, if the interior is purely passive but nonetheless
remains homogeneous, is
\[
\sigma _{\phi \phi }=\frac{\rho }{3l}\Omega ^{2}R^{3}\sin ^{3}\theta
\]
The density of the inner material is $\rho $, while $l$ is the thickness of
the membrane. There will be a contribution to the source tensor only at the
membrane.

The presence of the elastic tensor, even in the simplified version of a
surface tension, is necessary, from the mathematical point of view, to
insure the continuity of the radial derivatives of the metric tensor at the
boundary of the body. In fact from (\ref{linesex}) and (\ref{linesint}) the
derivatives of the metric at the boundary turn out to be (considering the $%
00 $ and $rr$ components) $\frac{2}{R}\beta $ on the internal side and $-%
\frac{2}{R}B_{0}$ on the external side. A $-\frac{4}{R}\beta $ term is
needed on the left to restore the equality. The additional surface term
corresponds to the elastic force originated by the elastic stress in the
membrane (per unit mass and dividing by $c^{2}$).

The stress affordable by the membrane has an upper limit $\sigma _{m}$, not
depending on the density of the membrane by itself. This puts an upper limit
to the angular velocity of the body too. The origin of the elastic
resistance is in molecular interactions, i.e. in electromagnetic
interaction. At the scale of the laboratory electromagnetic forces are far
greater than the gravitational force. This is why gravitational effects can
be neglected while rotation effects, whose limit is determined by elastic
stresses, can not. Increasing the size of the body the gravitational effect
keeps growing, while the molecular forces, which are short ranged, stay more
or less the same: this is the reason why for big bodies gravity overtakes
again and eventually rotational effects become secondary and more or less
negligible.

However it is important to remark that no zero mass limit can produce any
paradox since there are no electric charges without a mass and
relativistically the electromagnetic field also contributes to the mass. So
any zero mass limit is also a zero electromagnetic interaction limit.
Sending the mass to zero implies turning the elastic interaction off and the
weaker this is the smaller is the maximum possible angular velocity, then
the limit for $\rho $ going to zero is also an $\Omega $ going to zero limit.

\section{Experimental verification}

Let us refer to (\ref{linesex}) and suppose to send light along a closed
optical fiber at constant $r$ and constant $\theta $. The corresponding line
element would be
\begin{equation}
0=c^{2}\left( 1+\frac{8}{3}\pi \frac{\Omega ^{2}}{c^{2}}\frac{R^{4}}{r^{2}}%
\cos \theta \right) dt^{2}-r^{2}\sin ^{2}\theta \left( 1-\frac{8}{3}\pi
\frac{\Omega ^{2}}{c^{2}}\frac{R^{4}}{r^{2}}\cos \theta \right) d\phi ^{2}
\nonumber
\end{equation}
The coordinate time of flight along the whole loop would be
\begin{eqnarray*}
t_{1} &=&\frac{2\pi }{c}r\sin \theta \sqrt{\frac{\left( 1-\frac{8}{3}\pi
\frac{\Omega ^{2}}{c^{2}}\frac{R^{4}}{r^{2}}\cos \theta \right) }{\left( 1+%
\frac{8}{3}\pi \frac{\Omega ^{2}}{c^{2}}\frac{R^{4}}{r^{2}}\cos \theta
\right) }} \\
&\simeq &\frac{2\pi }{c}r\sin \theta \left( 1-\frac{8}{3}\pi \frac{\Omega
^{2}}{c^{2}}\frac{R^{4}}{r^{2}}\cos \theta \right)
\end{eqnarray*}

Let us now do the same using an identical loop at $\theta ^{\prime }=\pi
-\theta $. The time of flight would be
\[
t_{2}\simeq \frac{2\pi }{c}r\sin \theta \left( 1+\frac{8}{3}\pi \frac{\Omega
^{2}}{c^{2}}\frac{R^{4}}{r^{2}}\cos \theta \right)
\]
The difference between the two times is of course
\[
\Delta t=\frac{32}{3}\pi ^{2}\frac{\Omega ^{2}R^{4}}{c^{3}r}\sin \theta \cos
\theta
\]
This difference would not be there if the body was not rotating. The maximum
effect is produced when $\theta =\frac{\pi }{4}$.
\[
\Delta t_{M}=\frac{16}{3}\pi ^{2}\frac{\Omega ^{2}R^{4}}{c^{3}r}\simeq \frac{%
16}{3}\pi ^{2}\frac{\Omega ^{2}R^{2}}{c^{2}}\frac{R}{c}
\]
The value of $\Delta t_{M}$ in laboratory conditions ($R\sim 1$ m, $\Omega
R\sim 10^{3}$ m/s) is $\sim 10^{-18}$ s. It could be measured by
interferometric techniques.

\section{Conclusion}

Summarizing the results of this paper, we have shown that:

1) there exists a metric tensor in the vicinity of a weakly gravitating and
rotating body which can be expressed, at the chosen level of approximation,
as depending only on the square of the angular velocity of the body as
viewed by a static (with respect to the rotation axis) inertial observer and
on the shape of the body (here weakly gravitating means that the mass terms
are negligible with respect to rotation terms);

2) the Riemann tensor calculated from the external metric has non zero
terms, hence corresponding to physical and not only coordinate effects;

3) there exists an internal metric tensor matching the external one,
provided an elastic force is allowed to resist the centrifugal force within
the matter distribution;

4) the external metric tensor can be deduced from a source symmetric and
conserved (in the sense of null four-divergence) tensor, which turns out to
be proportional to the Einstein tensor calculated from the internal metric
tensor;

5) the relevant term in the source tensor as well as in the metric tensors
has the properties and appearance respectively of a dipolar density and
dipolar potential, whose magnitude is given by the square of the angular
velocity of the body;

6) considering the zero mass limit one sees that a vanishing mass
corresponds to a vanishing elastic force, then vanishing allowed angular
velocity too and eventually purely flat space time;

7) the numerical size of the various terms in ''laboratory'' conditions is
such that physical effects can be measured by light interferometry;

8) studying the structure of space time, rotation effects should be
accounted for adding to the energy momentum tensor a rotation source tensor.

\section{Acknowledgments}

The author wishes to thank the GPB group of the Stanford University for kind
hospitality and financial support during the elaboration of the present
paper and is particularly grateful to Alex Silbergleit for precious advise
and for spotting out the frequent computation mistakes which accompanied the
evolution of the work, and Francis Everitt, Ron Adler, Mac Keysen and Bob
Wagoner for many stimulating discussions.

\end{document}